\documentclass[twocolumn]{aastex631}

\newlength{\dzero}
\newcommand{\minnumb}{\settowidth{\dzero}{$-$}\kern-\dzero$-$}
\newcommand{\spcn}[1]{\settowidth{\dzero}{0}\kern#1\dzero}

\begin{document}

\title{Laboratory rotational spectroscopy leads to the first interstellar detection of singly deuterated methyl mercaptan (CH$_{2}$DSH)}

\author[0000-0002-7783-3049]{Hayley A. Bunn}
\affiliation{Center for Astrochemical Studies, Max-Planck-Institut für extraterrestrische Physik\\ Gießenbachstraße 1, 85748 Garching, Germany}

\author[0000-0002-6787-5245]{Silvia Spezzano}
\affiliation{Center for Astrochemical Studies,  Max-Planck-Institut für extraterrestrische Physik\\ Gießenbachstraße 1, 85748 Garching, Germany}

\author[0000-0003-0498-6957]{Laurent H. Coudert}
\affiliation{Université Paris-Saclay, CNRS, Institut des Sciences Moléculaires d’Orsay\\ 91405 Orsay, France}

\author[0000-0002-2929-057X]{Jean-Claude Guillemin}
\affiliation{Univ Rennes, Ecole Nationale Sup\'erieure de Chimie de Rennes, CNRS\\ ISCR – UMR6226, F-35000 Rennes, France}

\author{Yuxin Lin}
\affiliation{Center for Astrochemical Studies,  Max-Planck-Institut für extraterrestrische Physik\\ Gießenbachstraße 1, 85748 Garching, Germany}

\author{Christian P. Endres}
\affiliation{Center for Astrochemical Studies,  Max-Planck-Institut für extraterrestrische Physik\\ Gießenbachstraße 1, 85748 Garching, Germany}

\author{Brant Billinghurst}
\affiliation{Canadian Light Source. Inc., University of Saskatchewan\\ Saskatoon, Saskatchewan S7N 2 V3, Canada}

\author[0000-0002-4332-1440]{Olivier Pirali}
\affiliation{Université Paris-Saclay, CNRS, Institut des Sciences Moléculaires d’Orsay\\ 91405 Orsay, France}
\affiliation{AILES beamline Synchrotron Soleil, L’Orme des Merisiers, 91190 Saint-Aubin, France}

\author{Jes Jørgensen}
\affiliation{Niels Bohr Institute, University of Copenhagen, Øster Voldgade 5–7, 1350 Copenhagen K., Denmark}

\author{Valerio Lattanzi}
\affiliation{Center for Astrochemical Studies,  Max-Planck-Institut für extraterrestrische Physik\\ Gießenbachstraße 1, 85748 Garching, Germany}

\author{Paola Caselli}
\affiliation{Center for Astrochemical Studies,  Max-Planck-Institut für extraterrestrische Physik\\ Gießenbachstraße 1, 85748 Garching, Germany}

\begin{abstract}

We report an extensive rotational spectroscopic analysis of singly deuterated methyl mercaptan (CH$_{2}$DSH) using both millimeter and far-infrared synchrotron spectra to achieve a global torsional analysis of the three lowest torsional substates (\textit{e}$_{0}$, \textit{e}$_{1}$, and \textit{o}$_{1}$) of this non-rigid species. A fit including 3419 millimeter wave transitions along with 43 infrared torsional subband centers was performed with root mean square deviations of 0.233 MHz and 0.270 cm$^{-1}$, respectively, resulting in 68 fit parameters. A spectroscopic catalogue built from this analysis for a temperature of 125 K has led to the first interstellar detection of CH$_{2}$DSH towards the Solar-like protostar IRAS 16293-2422 B. We report the identification of 46 transitions, including eight relatively unblended lines, resulting in a derived column density of (3.0$\pm$0.3)$\times$10$^{14}$ cm$^{-2}$. The column density ratio for HDCS/CH$_{2}$DSH compared to HDCO/CH$_{2}$DOH suggests a difference in the interstellar chemistry between the sulphur and oxygen complex organics, in particular a different link between H$_{2}$CO and CH$_{3}$OH and between H$_{2}$CS and CH$_{3}$SH. This is the first interstellar detection of a deuterated sulphur-bearing COM and therefore an important step into understanding the chemical origin of sulphur-based prebiotics.

\end{abstract}

\section{Introduction} \label{sec:intro}

Sulphur is one of the most abundant elements in the interstellar medium (S/H $\sim$ 1.35×10$^{-5}$). Sulphur-bearing molecules are widely observed in star-forming regions to probe the physical properties and the chemistry of the gas (\citealt{ch3shism7, fuente23, mcclure23}). While being an essential component for the building blocks of life (\citealt{wu19, canavelli19}), the chemistry of sulphur and its prebiotic origin is one of the least understood in the field of astrochemistry. 

Isotopologue abundance ratios are pivotal for tracing the origin and evolution of the molecular material in the process of star and planetary system formation. In particular, deuterium fractionation has been shown as a useful tool for tracing a molecule's formation and evolution in low-mass star-forming regions (\citealt{deutfrac}). A common deuteration fraction is observed for oxygen-based complex organic molecules towards low mass protostars (e.g. acetaldehyde, \citealt{aa}, methyl formate, \citealt{mf}, and dimethyl ether, \citealt{dme}). Moreover, in \citealt{presetlmeth, me} a clear link has been shown between prestellar methanol and cometary methanol. The deuterated isotopologues of sulphur COMs, however, are yet to be unambiguously detected and therefore, little to nothing is known about the inheritance of sulphur into COMs. 

Methyl mercaptan, CH$_{3}$SH, the sulphur analog of methanol, is the simplest sulphur-based COM. CH$_{3}$SH was first observed in the interstellar medium (ISM) towards Sgr B2 (\citealt{tentch3shism, firstch3shism}) and has since been observed towards multiple star-forming regions (\citealt{ch3shism3,ch3shism5,ch3shism6,ch3shism9,pils,ch3shism,ch3shism2,ch3shism8,ch3shism7,ch3shism4}), with high enough abundances that make detection of its deuterated isotopologues plausible in some sources. The search for deuterated isotopologues of CH$_{3}$SH, however, is limited by the lack of spectroscopy on these molecules. So far, only the spectrum of monodeuterated methyl mercaptan, with the deuterium on the sulfur atom, CH$_{3}$SD, is known with an accuracy that can enable its search in the ISM (\citealt{ch3sdmw,ch3sd}). While the search for CH$_{3}$SD in star-forming regions was unsuccessful, the deuterated isotopologue most likely to be observed is expected to be CH$_{2}$DSH, as seen for methanol (e.g. \citealt{PILSoxygen}). The laboratory microwave spectra of CH$_{2}$DSH and CHD$_{2}$SH were first observed in 1970 (\citealt{1std1_quade}). The microwave spectrum has since been extended and ground and torsionally excited state transitions  characterised by very low \textit{J}- and \textit{K}$_{a}$-values have been assigned (\citealt{su1,su2}). Since the early works of \cite{su1}, significant efforts at refining the torsional potential and improving the microwave assignments have been made, providing well-determined rotational and torsional coupling parameters (\citealt{MUK97,pesch2dsh, assignch2dsh, quade07}). Despite the quality of the experimental microwave data, the frequency range covered in these early studies is still not sufficient for accurate extrapolation up to the higher frequencies necessary for astronomical observation. Therefore, additional spectral information is required to allow for its unambiguous interstellar detection. 

In this letter, we report the millimeter-wave and far-infrared spectra of singly deuterated methyl mercaptan, CH$_{2}$DSH, together with a line position analysis of these new measurements and a fit to a dedicated model. The spectroscopic database built from this analysis has allowed for the first detection of this species towards the protostellar region, IRAS 16293-2422 B. This study is, therefore, the first step towards revealing the details of how sulphur organics are formed and inherited into planets and prebiotics. 


\section*{Methods} \label{sec:exp}
\subsection*{Synthesis of CH$_{2}$DSH}
The synthesis was performed at the Institute of Chemical Sciences in Rennes, France and the final gas production step performed at the Max Planck Institute for Extraterrestrial Physics, Garching, Germany. The methods of this synthesis are outlined in Appendix A. The sample was obtained with higher than 95$\%$ isotopic purity. 

\subsection*{Millimeter Spectral Acquisition}
Spectra were first collected using a Chirped-Pulse Fourier Transform Spectrometer (CP-FTS, 75--110 and 150--220 GHz) for initial search and identification of CH$_{2}$DSH transitions with low quantum numbers and low frequencies, where the spectral prediction is most accurate. The spectra were then extended using the frequency modulated absorption experiment (82.5--1000 GHz), explained previously in \citealt{aa}.

The CP-FTS employed an arbitrary waveform generator (Keysight M8190A) to generate a 200 ns-long chirped pulse, which was initially upconverted to the 12--18 GHz frequency band through the application of an IQ mixer and a tunable signal generator acting as a local oscillator. Subsequently, it was amplified, and then fed into an active sixtupler to obtain frequencies in the range of 75--110 GHz. It was then further amplified (to a power of 250 mW) before entering the chamber \textit{via} a horn antenna. On the detection path, the molecular signal was first amplified with a low noise amplifier, downconverted to DC-2.5 GHz using a second tunable signal generator acting as a local oscillator, and finally recorded by an Acqiris U5310A digitiser card with a 2.5 GHz bandwidth. To extend the spectrometer to the 2 mm wavelength range, a frequency doubler was incorporated between the amplifier and the horn antenna in the signal generation path, and a subharmonic mixer was introduced in place of the first low-noise amplifier in the detection path. The 3 mm and 2 mm wavelength ranges were covered by tuning the local oscillators in steps of 1.8 GHz and recording 20 and 40 spectra, respectively. In total, 4 million and 16 million acquisitions were recorded for each spectrum, respectively, at a rate of 10,000 acquisitions per second. The frequency spectra were finally obtained by Fast Fourier Transform (FFT) of the averaged time-domain spectra and show a resolution of 76 kHz.

The frequency modulated absorption spectrometer was composed of a microwave synthesizer (Keysight E8257D) and Virginia Diodes Inc. frequency multipliers coupled to various doublers and triplers to access frequencies between 82.5--1000 GHz. The signal is frequency modulated with a sine wave at a rate of 15 kHz and a width of 250 kHz. The signal is detected using an InSb Hot electron bolometer (QMC Ltd. XBI with a closed-cycle He compressor) and processed through a lock-in amplifier (SR830, Stanford Research Systems) using a 3 ms time constant and 2f detection. The spectra were collected at a step size of 30 kHz with 2 averages.

In both cases the gaseous sample was introduced into a 2 m long stainless steel cell held at a static pressure of 10--20 $\mu$Torr.  

\subsection*{Far-infrared Spectral Acquisition}
The torsional subbands of CH$_{2}$DSH, located in the far-infrared region, were observed using facilities at both the Canadian Light Source (CLS) and the French National Synchrotron Facility (SOLEIL). The frequency region from 100--600 cm$^{-1}$ was collected using the far-infrared beamline at CLS utilising full resolution capabilities (0.00096 cm$^{-1}$) of the Bruker IFS 125HR FTIR. These frequencies were accessed using a 6 $\mu$m Mylar beamsplitter and QMC Superconducting Niobium TES Bolometer. The spectra were recorded using a White multipass cell with a 72 m pathlength, equipped with polypropylene windows. The sample spectra were collected at a static sample pressure of 1.045 Torr and averaged 317 times for optimal signal to noise ratio of the torsional subbands. The AILES beamline at SOLEIL were used to extend the spectra down to 12 cm$^{-1}$. All measurements were performed using a White multipass cell equipped with 50 $\mu$m polypropylene windows and set to a 100 m path length. Two different detectors were needed to achieve the required sensitivity and coverage below 100 cm$^{-1}$. For optimal sensitivity between $\sim$ 12--60 cm$^{-1}$ a 1.6 K pumped InSb bolometer along with a 50 $\mu$m Mylar beamsplitter were used. A 4.2 K silicon bolometer was then used for the higher frequency portion. The background scans were recorded at 0.02 cm$^{-1}$ resolution, the lowest resolution required to resolve the fringes. The sample scans were performed using static pressures between 0.2 and 3.5 mbar to observe both intense and weak transitions. The scans were recorded at the full resolution capabilities of the Bruker IFS 125HR FTIR (0.00102 cm$^{-1}$), collecting $\sim$100 averages. \\


\section{Results} \label{sec:style}

\subsection*{Spectral Analysis}
\begin{figure*}
      \centering
      \begin{minipage}[c]{0.4\textwidth}
        \includegraphics[width=10.4cm]{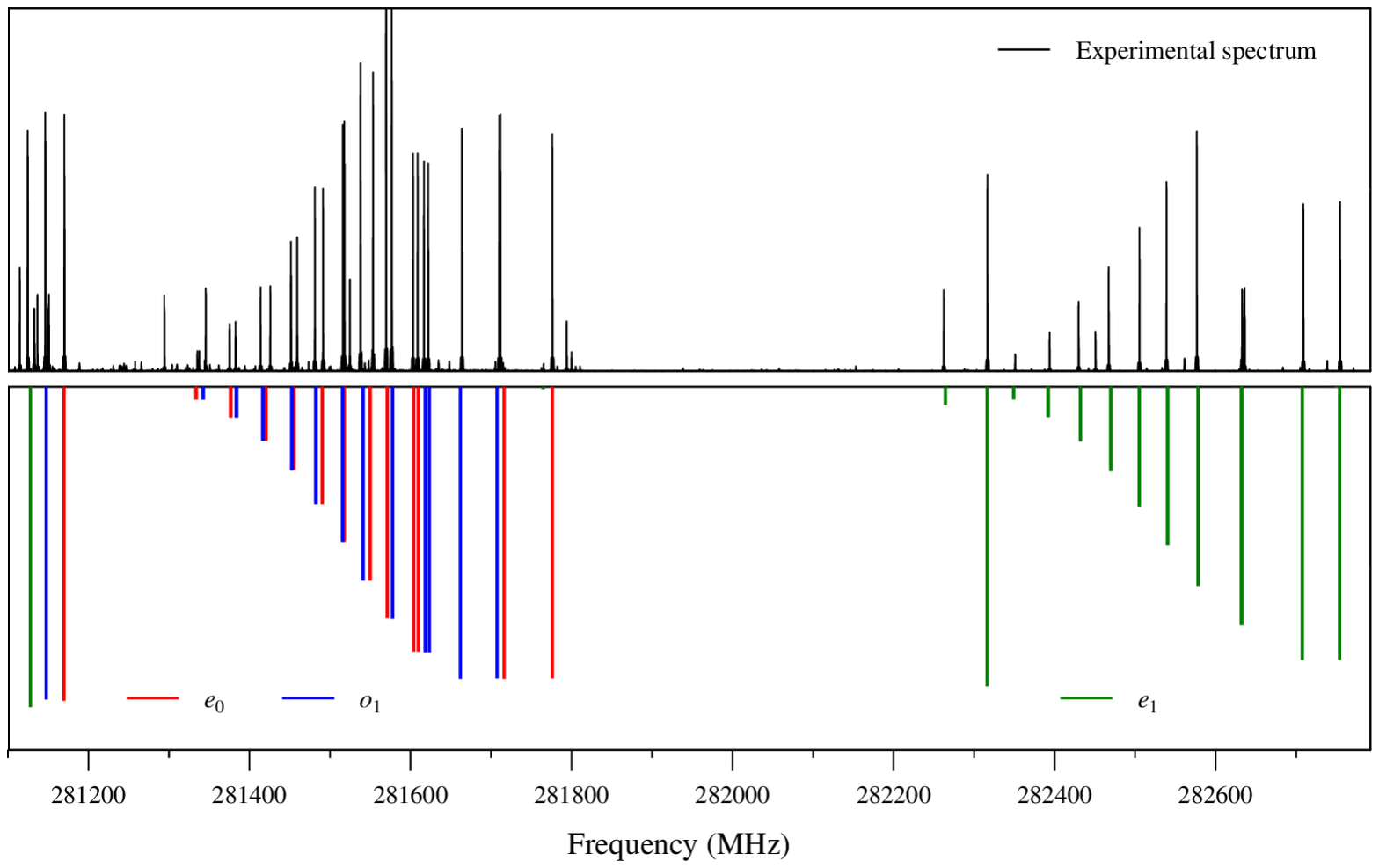}
      \end{minipage}
      \hfill
      \raisebox{3ex}
      {
      \begin{minipage}[c]{0.38\textwidth}
        \includegraphics[width=5.8cm]{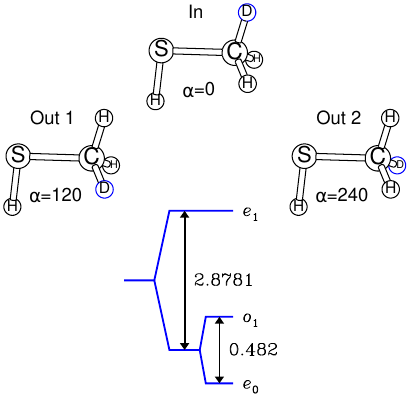}
      \end{minipage}
      }
    \caption{Left panel, a comparison between the experimental rotational spectrum (top) and calculated stick spectrum (bottom) in the region between 281.1 and 282.8 GHz of the \textit{$^{a}$}R$_{0,1}$(11) transitions. A prominent \textit{K}$_{a}$- and $\nu_{t}$-structure can be seen. Right panel, the three nonsuperimposable equilibrium configurations of CH$_{2}$DSH labeled Out 1, Out 2, and In, are shown. The equilibrium value of the torsional angle $\alpha = \pi + \angle$HSCD is given for each configuration. The \textit{J} = 0 energy level diagram is also shown and consists of three torsional states labeled \textit{e$_{0}$}, \textit{o$_{1}$}, and \textit{e$_{1}$} corresponding respectively to $\nu_{t}$=0, 1, and 2.}
    \label{arband}
    \end{figure*}

Methyl mercaptan (CH$_{3}$SH) is a near-prolate asymmetric-top with $\kappa$ = $-$0.988, as deduced from rotational constants reported by \citealt{ch3sh}, and dipole moment components $\mu_a$ = 1.312 D and $\mu_b$ = $-$0.758 D (\citealt{ch3sdmw}). It is a non-rigid molecule displaying internal rotation of its methyl group with three isoenergetic equilibrium configurations displaying \textit{C}$_{s}$ symmetry. By analogy with CH$_{3}$SH, CH$_{2}$DSH also undergoes internal rotation, however, because of the introduction of the deuterium atom, the torsional potential no longer has threefold symmetry. There arise two isoenergetic \textit{C}$_{1}$ symmetry minima, labeled Out 1 and Out 2 in Fig.~\ref{arband}, with the deuterium atom outside the CSH plane and a higher lying \textit{C}$_{s}$ symmetry minimum, labeled In in Fig.~\ref{arband}, with the deuterium atom in the symmetry plane. The three lowest lying torsional levels are shown in Fig.~\ref{arband} and consist of the close lying \textit{e$_{0}$} and \textit{o$_{1}$} levels arising from tunneling between the Out 1 and Out 2 configurations and the higher lying isolated \textit{e$_{1}$} level corresponding to the In configuration. The rotation-torsion coupling leads to multiple interactions within and between these states so that their rotational energies differ from those of a rigid rotator. These rotation-torsion levels can be labeled with the rotational quantum numbers \textit{J} and \textit{K}$_{a}$, and the torsional quantum number $\nu_{t}$.

The spectra of CH$_{2}$DSH can be described in terms of the torsional subbands \textit{K}$_{a}'$, $\nu_{t}'$ $\leftarrow$ \textit{K}$_{a}''$, $\nu_{t}''$ characterized by the rotational and torsional quantum numbers \textit{K}$_{a}$ and $\nu_{t}$ of the upper and lower torsional states. When $\Delta K_{a} = \Delta\nu_{t} = 0$, we have strong parallel \textit{a}-type intrastate lines which dominate the low frequency region of the spectrum. Perpendicular subbands with $\Delta K_{a} = \pm 1$ and $\Delta\nu_{t} = 0$ display \textit{R-, Q-,} and \textit{P-}branches consisting of weaker \textit{b}- or \textit{c}-type interstate lines. Perpendicular subbands with $\Delta K_{a} = \pm 1$ and $\Delta\nu_{t}>0$, tend to be even weaker and give rise to lines in the far-infrared region. Figure \ref{arband} shows parallel \textit{$^{a}$}R$_{0,1}$(11) transitions where prominent \textit{K}$_{a}$- and $\nu_{t}$-structures can be seen. The low frequency transitions involve the $\nu_{t}$ = 0 and 1 torsional states; the high frequency transitions involve the $\nu_{t}$ = 2 torsional state characterized by different rotational constants than the previous ones. 

\begin{figure*}[ht!]
\centering
	\includegraphics[scale=0.45]{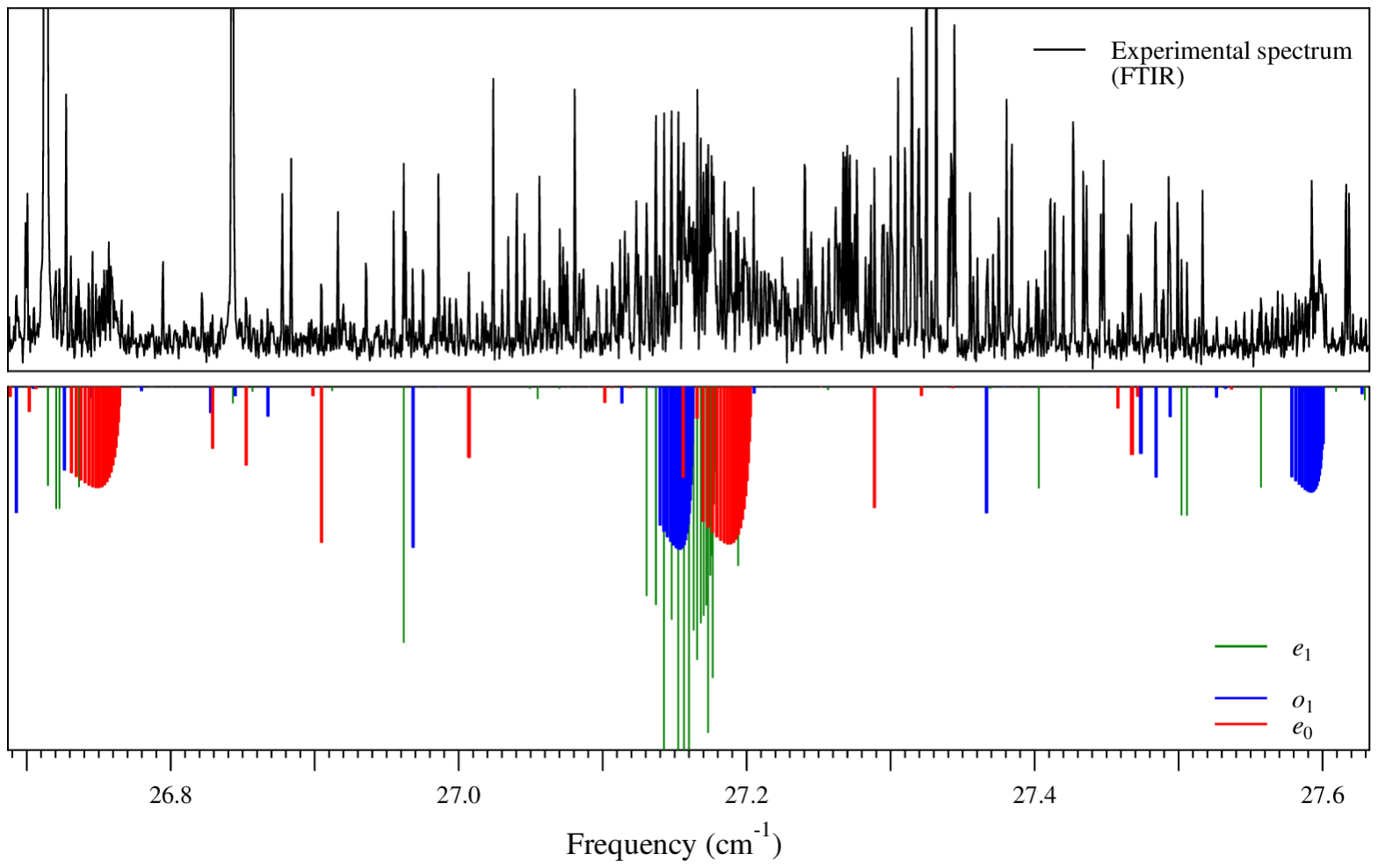}
	\caption{The experimental far-infrared spectrum (top) of CH$_{2}$DSH showing the Q-branch of three intrastate ($\Delta\nu_{t}$=0) torsional subbands, near 27.2 cm$^{-1}$, and two interstate ($\Delta\nu_{t}$=$\pm$1) torsional subbands near 26.75 cm$^{-1}$ and 27.6 cm$^{-1}$. The calculated spectrum (bottom) emphasizes the strong overlap between bands, especially near 27.2 cm$^{-1}$. Calculated transitions are colour coded with red, blue, and green indicating that their lower torsional state is \textit{e$_{0}$}, \textit{o$_{1}$}, and \textit{e$_{1}$}, respectively.}
	\label{ir}
\end{figure*}

Initial attempts at fitting this spectra using the approaches described in \cite{mfMW} and \cite{fish} allowed for reasonable assignments and predictions for the \textit{$^{a}$}R$_{0,1}$ parallel transitions with \textit{K$_{a}$}$\ge$3 as well as assignment of the weaker \textit{b}-type Q-branch transitions (\textit{$^{b}$}Q$_{1,-1}$). For \textit{K}$_{a}<3$, the increased severity of the interactions between the three lowest lying torsional states precludes the use of a Watson-type Hamiltonian, even for the isolated \textit{e$_{1}$} state. The high-barrier effective approach developed for methyl formate (\citealt{mfMW}) and successfully applied to mono- and bi-deuterated acetaldehyde (\citealt{coudert19, aa}), which incorporates the couplings between the three torsional states, only allowed us to model low \textit{J}-value transitions. As a result, the global analysis method described previously (\citealt{globalanalysis}) has been used. The use of this approach in the present work, however, requires accurate determination of the torsional potential, which, as in the case of CH$_{2}$DOH (\citealt{Haliali}), was obtained from a line position analysis of the subband centers of torsional subbands assigned in the far-infrared spectrum ($7 \leq K_{a} \leq 14$). Their rotational structure could not be analyzed due to strong overlap and their Q-branch wavenumber was taken as their subband center (see Fig.~\ref{ir}). The rotational structure of the torsional subbands up to \textit{K}$_{a}=7$ could, however, be assigned in the millimeter spectrum and are also important to determine the torsional potential. These Q-branch torsional bands with $\Delta\nu_{t} = 0$ were usually observed, although weaker than the $\Delta K_{a} = \Delta\nu_{t} = 0$ \textit{a}-type R-branch lines. Fewer torsional subbands with $\Delta\nu_t >0$ were observed as they are even weaker. Almost no $\Delta\nu_t >0$ transitions belonging to torsional subbands involving the $\nu_{t}=2$ (\textit{e$_{1}$}) torsional state were observed. This can be ascribed to the higher torsional barrier in CH$_{2}$DSH than in CH$_2$DOH ($V_{3}\approx$ 440 vs 373 cm$^{-1}$), where many such transitions could be observed.

The fitted data set consists of 3419 millimeter transitions and 43 infrared torsional subband centers. The frequency of the measured millimeter/submillimeter lines were given an experimental uncertainty of 0.06 MHz; for the infrared lines, an experimental uncertainty of 0.1 cm$^{-1}$ was used. In the least-squares fit procedure, all lines were given a weight equal to the inverse of their experimental uncertainty squared. The millimeter transitions involve all three torsional substates and include parallel and perpendicular transitions. These are either intrastate ($\Delta\nu_{t}$=0) or interstate ($\Delta\nu_{t}$=$\pm$1) lines. The maximum \textit{J} and \textit{K$_{a}$} values are 24 and 12, respectively, and 68 rotation-torsion parameters, presented in Table \ref{prm_table}, were determined. The resulting fit presented here has a root mean square (RMS) deviation for the observed minus calculated residuals of 0.27 MHz and 0.233 cm$^{-1}$ for millimeter and infrared transitions, respectively. These RMS values compare well to the 0.21 MHz reported for monodeuterated methanol (\citealt{globalanalysis}), which resulted from 7445 millimeter transitions, 668 far-infrared transitions, 98 torsional subband centers, and 103 parameters. The unitless standard deviation for the fit is 4.5, which is almost double what is observed for the methanol fit, 2.4. The higher standard deviation and RMS are due to the larger effects of the rotation-torsion couplings in CH$_{2}$DSH than in CH$_{2}$DOH because the former displays smaller torsional splittings than the latter. While efforts to reduce this error, include more transitions, and further improve the full spectral analysis is on-going, the spectral prediction up to \textit{J}=25 was produced based on this fit providing reliable frequencies for the intense transitions required for interstellar analysis. This information is sufficient to allow for a confident search for this species in the ISM, \textit{vide infra}. The spectral catalog described here will be made available through the Cologne Database for Molecular Spectroscopy (CDMS: \href{http://www.cdms.de}{http://www.cdms.de}, \citealt{cdms05}).

\begin{table*}
\caption{\label{prm_table}Rotation-torsion spectroscopic parameters resulting from the global analysis of the millimeter and far-infrared spectra of CH$_{2}$DSH.}
\begin{small}
\begin{tabular}{@{}ll@{\hspace*{0.5\tabcolsep}}l@{\hspace*{2.6\tabcolsep}}
                   ll@{\hspace*{0.5\tabcolsep}}l@{\hspace*{2.6\tabcolsep}}
                   ll@{\hspace*{0.5\tabcolsep}}l@{}} \\ \hline\hline
\makebox[3.0em][l]{Parameter$^a$} & \multicolumn{2}{c}{Value$^b$} &
\makebox[3.0em][l]{Parameter$^a$} & \multicolumn{2}{c}{Value$^b$} &
\makebox[3.0em][l]{Parameter$^a$} & \multicolumn{2}{c}{Value$^b$} \\ \hline
$I_{x}^0$ & 4.0302(11) & $\times 10^{1}$  &
$l_k$ & 1.953(33) & $\times 10^{-5}$  &
$H_{J}$ & 1.502(85) & $\times 10^{-11}$  \\
$I_{y}^0$ & 4.3025(14) & $\times 10^{1}$  &
$k_{2k}$ & \minnumb1.751(26) & $\times 10^{-5}$  &
$h_{K}$ & \minnumb1.19(27) & $\times 10^{-8}$  \\
$I_{z}^0$ & 5.94113(9) &  &
$k_{4B}$ & 4.534(34) & $\times 10^{-6}$  &
$h_{K\!J}$ & \minnumb7.77(84) & $\times 10^{-10}$  \\
$I_{xz}^0$ & 9.5122$^{c}$ & $\times 10^{-1}$  &
$V_{1J\!J}$ & 5.929(18) & $\times 10^{-3}$  &
$h_{J}$ & \minnumb3.14(36) & $\times 10^{-12}$  \\
$I_{x}^1$ & \minnumb3.252(11) & $\times 10^{-2}$  &
$V_{2J\!J}$ & 2.529(15) & $\times 10^{-3}$  &
$G_2$ & \minnumb1.60(14) & $\times 10^{-3}$  \\
$I_{\alpha}$ & 8.00(14) & $\times 10^{-1}$  &
$V_{3J\!J}$ & \minnumb1.7514(69) & $\times 10^{-2}$  &
$G_3$ & 2.43(310) & $\times 10^{-4}$  \\
$I_x$ & \minnumb1.1867(35) &  &
$V_{4J\!J}$ & \minnumb1.3638(23) & $\times 10^{-2}$  &
$G_4$ & \minnumb4.39(53) & $\times 10^{-4}$  \\
$I_{\alpha}^0$ & 4.2099$^{c}$ &  &
$V_{5J\!J}$ & \minnumb2.297(11) & $\times 10^{-3}$  &
$G_5$ & 5.85(95) & $\times 10^{-4}$  \\
$I_{xz J\!J}^0$ & \minnumb7.16(37) & $\times 10^{-5}$  &
$V_{6J\!J}$ & 1.981(17) & $\times 10^{-3}$  &
$G_6$ & \minnumb5.98(280) & $\times 10^{-4}$  \\
$I_{x J\!J}^1$ & 3.74(26) & $\times 10^{-6}$  &
$\Delta_{xz}$ & 4.56(20) & $\times 10^{-4}$  &
$V_{1K\!K\!J\!J}$ & 9.08(330) & $\times 10^{-7}$  \\
$I_{\alpha J\!J}$ & \minnumb6.77(80) & $\times 10^{-6}$  &
$H_1$ & 1.5406(96) & $\times 10^{-2}$  &
$V_{2K\!K\!J\!J}$ & \minnumb1.88(120) & $\times 10^{-7}$  \\
$I_{x J\!J}$ & \minnumb2.78(28) & $\times 10^{-5}$  &
$\delta_{xz}$ & \minnumb4.01(18) & $\times 10^{-4}$  &
$V_{3K\!K\!J\!J}$ & \minnumb2.20(29) & $\times 10^{-6}$  \\
$V_1$ & 4.7003(46) &  &
$\rho_{xJ}$ & \minnumb1.293(23) & $\times 10^{-5}$  &
$V_{1J\!J\!J\!J}$ & \minnumb1.88(200) & $\times 10^{-8}$  \\
$V_2$ & \minnumb9.6386(77) &  &
$L_v$ & 9.821(60) & $\times 10^{-4}$  &
$V_{2J\!J\!J\!J}$ & 1.71(22) & $\times 10^{-7}$  \\
$V_3$ & 4.42328(31) & $\times 10^{2}$  &
$G_v$ & \minnumb7.019(40) & $\times 10^{-4}$  &
$V_{3J\!J\!J\!J}$ & 7.21(32) & $\times 10^{-7}$  \\
$D_K$ & 1.119(18) & $\times 10^{-4}$  &
$G_1$ & \minnumb1.12(33) & $\times 10^{-4}$  &
$V_{4J\!J\!J\!J}$ & \minnumb2.29(32) & $\times 10^{-7}$  \\
$D_{K\!J}$ & 3.380(24) & $\times 10^{-4}$  &
$c_1$ & \minnumb5.80(510) & $\times 10^{-6}$  &
$V_{5J\!J\!J\!J}$ & 8.64(240) & $\times 10^{-8}$  \\
$D_J$ & \minnumb1.61(27) & $\times 10^{-7}$  &
$c_4$ & \minnumb1.99(87) & $\times 10^{-5}$  &
$V_{6J\!J\!J\!J}$ & \minnumb7.54(35) & $\times 10^{-7}$  \\
$d_k$ & \minnumb2.06(35) & $\times 10^{-5}$  &
$D_{1xy}$ & 1.029(42) & $\times 10^{-3}$  &
$V_{1K\!K\!K\!K}$ & \minnumb2.76(76) & $\times 10^{-6}$  \\
$d_j$ & \minnumb2.92(140) & $\times 10^{-9}$  &
$D_{1yz}$ & \minnumb3.864(49) & $\times 10^{-2}$  &
$V_{2K\!K\!K\!K}$ & \minnumb4.00(68) & $\times 10^{-6}$  \\
$k_2$ & 1.881(25) & $\times 10^{-4}$  &
$\rho_{y\alpha}$ & 3.572(46) & $\times 10^{-2}$  &
$V_{3K\!K\!K\!K}$ & \minnumb1.17(45) & $\times 10^{-5}$  \\
$V_{1K\!K}$ & \minnumb6.575(78) & $\times 10^{-3}$  &
$H_{K}$ & \minnumb5.62(11) & $\times 10^{-6}$  &
$V_{2K\!K\alpha\alpha}$ & \minnumb6.47(100) & $\times 10^{-4}$  \\
$V_{2K\!K}$ & 1.026(14) & $\times 10^{-2}$  &
$H_{K\!J}$ & 6.47(61) & $\times 10^{-9}$  &
\multicolumn{3}{c}{\mbox{}} \\
$V_{3K\!K}$ & 1.5271(71) & $\times 10^{-2}$  &
$H_{J\!K}$ & \minnumb5.06(45) & $\times 10^{-10}$  &
\multicolumn{3}{c}{\mbox{}}
 \\ \hline\hline
\multicolumn{9}{@{}l}{$^a$\parbox[t]{0.7\textwidth}{ Parameter constants described in \cite{globalanalysis}.}}\\
\multicolumn{9}{@{}l}{$^b$\parbox[t]{0.7\textwidth}{%
Generalized inertia tensor parameters $I$, involved in the Hamiltonian $H_0$ in Eq.~(1) of \cite{globalanalysis}, are given in amu \AA$^2$. All other parameters are given in cm$^{-1}$. Uncertainties are given in the same units as the last quoted digit.}}\\
\multicolumn{9}{@{}l}{$^c$\parbox[t]{0.7\textwidth}{Constrained to the value from the structure of CH$_{3}$SH (\citealt{xu12}).}}
\end{tabular}
\end{small}
\end{table*}

\subsection*{Astronomical Observations} \label{sec:astro}
Based on the new spectral information presented above, CH$_{2}$DSH was searched for towards the Solar-like protostar, IRAS 16293-2422 B, using the observed spectroscopic data from the ALMA PILS survey (\citealt{pils}). This survey is a wide-band unbiased line survey covering the frequency range 329.1--362.9 GHz. The velocity resolution is $\sim$0.2 km s$^{-1}$ with an angular resolution of 0.5$''$ (70 au). We adopted the observed spectrum at the offset position (0.5$''$ offset from the continuum peak of IRAS 16293-2422), the position where CH$_{3}$SH is detected and the parameters determined in \citet{PILSsulphur}.

The search for CH$_{2}$DSH was conducted whilst also including the other already detected species in the survey (\citealt{PILSoxygen}, \citealt{Coutens22}). Eight clear transitions of CH$_{2}$DSH, relatively free from contamination of nearby transitions of other known species (judging by the almost single gaussian profile), are identified (Fig. \ref{astro}). Additional blended transitions are identified and are provided in Appendix B. The frequencies, quantum numbers, upper state energies, degeneracies, and Einstein A coefficients are provided for all observed transitions in Table \ref{QNs}, and Fig. \ref{astrosi} shows the model comparison with contaminated transitions and weak lines. These contaminated transitions, however, do not have a significant impact on the detection or derivation of the column densities. An LTE model and a Markov Chain Monte Carlo method were used to fit and derive the column density of CH$_{2}$DSH, $N_{\mathrm{CH_{2}DSH}}$, adopting the constants from the newly generated catalog. We use a full width half-maximum (FWHM) of 1 km s$^{-1}$ and a temperature of 125 K, based on the excitation temperature of CH$_{3}$SH in this source (\citealt{PILSsulphur}). With a beam dilution factor of 0.5, and considering all 46 identified transitions a $N_{\mathrm{CH_{2}DSH}}$ of (3.0$\pm$0.3$)\times$10$^{14}$ cm$^{-2}$ is determined. When including only the 8 mostly unblended transitions shown in Figure \ref{astro} we derive a $N_{\mathrm{CH_{2}DSH}}$ of (3.5$\pm$1.0$)\times$10$^{14}$ cm$^{-2}$. Given that the column density when using the full set of blended and unblended transitions compared to just the 8 least blended are consistent, the following ratios are then reported using the more precise value of (3.0$\pm$0.3$)\times$10$^{14}$ cm$^{-2}$. The fitted model and 95$\%$ confidence level of the model is shown in Fig. \ref{astro}, and in Fig. \ref{astrosi}. For the blended CH$_{2}$DSH transitions, we ensure that the fitted model does not overpredict the line intensities, Fig. \ref{astrosi}. 

\begin{figure*}
    \centering
	\includegraphics[scale=0.55]{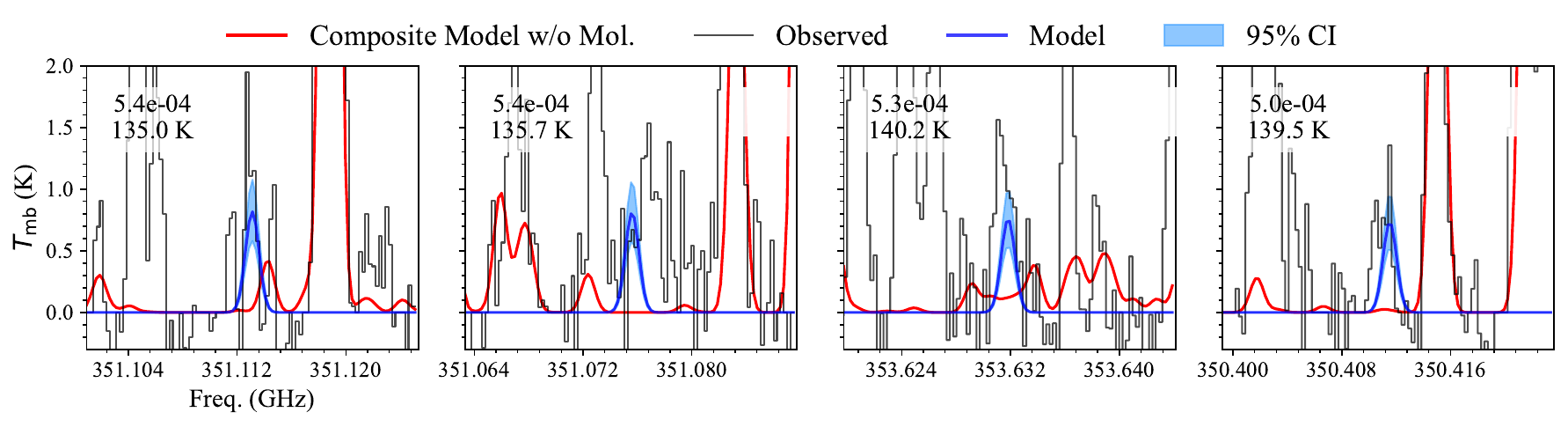}
 \centering 
 \includegraphics[scale=0.55]{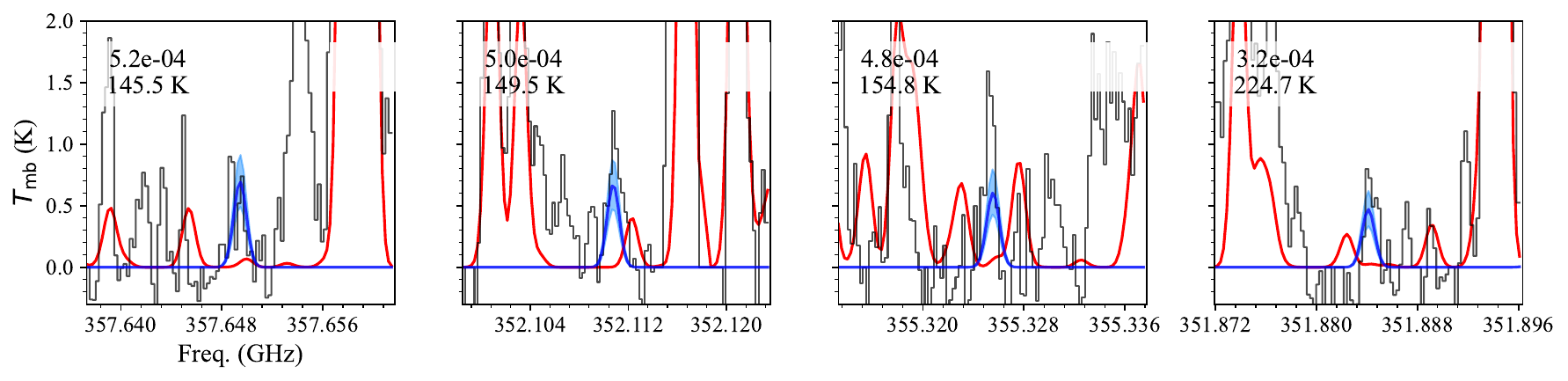}
	\caption{Eight detected unblended transitions of CH$_{2}$DSH in IRAS 16293-2422 B, where the black trace is the PILS data, the red is the fit of other known molecules and the blue is the fit to CH$_{2}$DSH. The top left of each panel displays the Einstein A coefficient (in s$^{-1}$) and upper energy level (in K).}
	\label{astro}
\end{figure*}

\begin{table*}
\small
	\caption{Column density ratios of CH$_{2}$DSH compared with H$_{2}$CS, CH$_{3}$OH, and H$_{2}$CO towards IRAS 16293-2422 B. The superscripted letters indicate the reference for the column density in IRAS 16293-2422 B and the roman numeral in L1544.}
	\centering
        \setlength{\tabcolsep}{1pt}
	\begin{tabular}{wl{1.8in}lwl{0.6in}l}
		 \\
		\hline\hline
        Ratio  & &Value (\%)&\\
         &  IRAS 16293-2422 B && L1544\\
		\hline
  CH$_{2}$DSH$^{a}$/CH$_{3}$SH$^{b}$ & 5.5 $\pm$ 0.5 && N/A \\
  HDCS$^{b,\rm I}$/H$_{2}$CS$^{b,\rm I}$ & 10.0 $\pm$ 1.4 && 11.6 $\pm$ 1.8 \\
  H$_{2}$CS$^{b,\rm I}$/CH$_{3}$SH$^{b,\rm II}$ & 27.3 $\pm$ 7.71 && 2760.0 (lower limit) \\
  HDCS$^{b,\rm I}$/CH$_{2}$DSH$^{a}$ & 50 $\pm$ 11 && N/A \\ \\
  
  CH$_{2}$DOH$^{c,\rm III}$/CH$_{3}$OH$^{c,\rm IV}$ & 7.1 $\pm$ 2.0 && 21.5 $\pm$ 5.4 \\
  HDCO$^{d,\rm I}$/H$_{2}$CO$^{d,\rm I}$ & 6.8 $\pm$ 1.2 && 3.5 $\pm$ 2.2 \\
  H$_{2}$CO$^{d,\rm I}$/CH$_{3}$OH$^{c,\rm IV}$ & 19.0 $\pm$ 5.4 && 282.3 $\pm$ 177.2 \\
  HDCO$^{d,\rm I}$/CH$_{2}$DOH$^{c,\rm III}$ & 18.3 $\pm$ 5.18 && 46.4 $\pm$ 12 \\
		\hline \hline
  \multicolumn{2}{l}{$^{a}$ Column density derived from the current work.} \\
  \multicolumn{2}{l}{$^{b}$ \citealt{PILSsulphur}.} \\
  \multicolumn{2}{l}{$^{c}$ \citealt{PILSoxygen}.} \\
  \multicolumn{2}{l}{$^{d}$ \citealt{formald}.} \\
  \multicolumn{2}{l}{$^{\rm I}$ \citealt{spezz22}.} \\
  \multicolumn{2}{l}{$^{\rm II}$ \citealt{ch3shism7}.} \\
  \multicolumn{2}{l}{$^{\rm III}$ \citealt{chacon19}.}\\
  \multicolumn{2}{l}{$^{\rm IV}$ \citealt{lin22}.}
		\label{ratio}
	\end{tabular}
\end{table*}

\section*{Discussion of astronomical implications}
Table \ref{ratio} presents the column density ratios involving H$_{2}$CS, CH$_{3}$SH, their singly deuterated isotopologues, and ratios for the equivalent oxygen system from observations towards the IRAS 16293-2422 B protostar and the prestellar core, L1544. Based on the column density derived here for CH$_{2}$DSH, we see that the ratio for CH$_{2}$DSH/CH$_{3}$SH (5.5$\pm$0.5\%) and CH$_{2}$DOH/CH$_{3}$OH (7.1$\pm$2\%) agree within error bars. The observed CH$_{2}$DOH/CH$_{3}$OH ratios towards pre-stellar cores and hot corinos range between 0.01 to 0.1 (See Figure 2 of \citealt{lin23} and references therein). The observed D/H from CH$_{2}$DOH/CH$_{3}$OH compared with current theoretical studies of methanol deuteration based on gas-grain chemical models (\citealt{taquet14,caselli02,riedel03}) and laboratory measurements (\citealt{nagaoka05, reactmeth}) seems to be consistent. While the deuteration of methanol has been the topic of a number of astronomical  investigations (e.g. \citealt{presetlmeth,lin23,parise02,parise04,bizzocchi14}), this is the first observation of CH$_{2}$DSH and therefore additional similar investigations into deuterated methyl mercaptan towards IRAS 6293-2422 B and other sources are required. Given the much lower abundance of CH$_{3}$SH in prestellar cores, however, it is not feasible to observe its deuterated isotopologue. For example, in L1544, while H$_{2}$CS is observed, only an upper limit for CH$_{3}$SH has been reported (\citealt{ ch3shism7}). Additionally, comparison of the deuteration within each molecule is also an important clue towards disentangling the chemistry of these species (\citealt{spezz22}). The ratio of CH$_{3}$OD/CH$_{2}$DOH is about 0.25 (\citealt{PILSoxygen}), however, CH$_{3}$SD has not yet been detected in the ISM and only an upper limit of $<$(8.8$\pm$0.9)$\times$10$^{14}$ cm$^{-2}$ towards IRAS16293-2422 B has been reported, resulting in CH$_{3}$SD/CH$_{2}$DSH  $<$3 (\citealt{ch3sd}). It is therefore not possible to draw any conclusion and further highlights the necessity of additional studies of deuterated CH$_{3}$SH towards IRAS 6293-2422 B and other sources, along with chemical models.

Previous studies into the deuteration of thioformaldehyde towards IRAS 16293-2422 B (\citealt{PILSsulphur}) report a HDCS/H$_{2}$CS column density ratio of 10$\pm$1.4\%, showing a higher deuteration level compared to formaldehyde, with a HDCO/H$_{2}$CO column density ratio of 6.8$\pm$1.2\% (\citealt{formald}). This result leads to the suggestion that this species may either form in more D-rich environments than H$_{2}$CO or undergo fewer chemical reactions that would result in a lower level of deuteration. As mentioned above, the column density ratios for CH$_{2}$DSH/CH$_{3}$SH and CH$_{2}$DOH/CH$_{3}$OH agree within errors bars. Similarly, the column density ratio for H$_{2}$CS/CH$_{3}$SH (27.3$\pm$7.71\%) and H$_{2}$CO/CH$_{3}$OH (19.0$\pm$5.4\%) are also within error bars. The ratio, however, for HDCX/CH$_{2}$DXH, where X=O or S, is significantly different, with the sulphur system showing a ratio over double that of the oxygen system, potentially suggesting a different chemical link between H$_{2}$CX/CH$_{3}$XH for sulphur and oxygen species. We know from previous studies (e.g., Figure 6 of \citealt{sanz} and references therein) that the O/S ratio for both H$_{2}$CX and CH$_{3}$XH show a significant variation depending on the environment and therefore further investigation into deuteration of these species in different regions such as clouds, cores, and comets will help towards discerning the chemical and physical process involving them. 

Based on the attempted observations of CH$_{3}$SH (\citealt{spezz22, ch3shism7}) the lower limit of the column density ratio, H$_{2}$CS/CH$_{3}$SH, is an order of magnitude higher than H$_{2}$CO/CH$_{3}$OH (\citealt{spezz22, lin22}), potentially indicating that the formation of CH$_{3}$SH from H$_{2}$CS is less efficient or that CH$_{3}$SH is still locked in the ice grains. This suggestion, however, is heavily dependent on many assumptions on the formation of CH$_{3}$SH. It is not well known how CH$_{3}$SH forms and what link it has to H$_{2}$CS, if it proceeds \textit{via} hydrogen atom addition to CS in ice mantels, like for methanol with CO, or if it involves reaction with H$_{2}$S and/or with H$_{2}$. While the deuteration of H$_{2}$CS has been studied rather extensively in L1544 (\citealt{spezz22}), the models are still unable to reproduce the observed abundances of D$_{2}$CO and D$_{2}$CS, further highlighting how little is known, particularly, in respect to the formation of H$_{2}$CS and CH$_{3}$SH. Therefore, in order to further constrain the chemistry of this system additional lab work similar to that done for methanol (e.g., \citealt{reactmeth}) is required.

\section*{Summary and Conclusions}

Deuterium fractionation is an important tool for tracing the chemical history and inheritance of molecules throughout
star and planet formation. In order to correctly determine the formation and inheritance of complex organic molecules (COMs) we, therefore, must also identify the associated isotopologues and their relative abundance. Despite being one of the most abundant elements, still so little is known about sulphur chemistry in the interstellar medium. In this work we present the first detection of a deuterated sulphur-bearing complex organic molecule, CH$_{2}$DSH, resulting from an extensive spectroscopic investigation, including both millimeter and infrared synchrotron measurements, also provided here. The spectra, however, are complicated by the hindered torsional rotation of the CH$_{2}$D group. The Hamiltonian needed to account for this unsymmetrical methyl rotor requires an accurate torsional potential, which is extracted from the torsional subbands present in the far-infrared, observed using the high-resolution far-infrared facilities at the CLS and SOLEIL. The higher column density ratio observed here for HDCS/CH$_{2}$DSH compared to HDCO/CH$_{2}$DOH in IRAS 16293-2422 B indicates that CH$_{3}$SH may have a different chemical pathway and/or form in a different environment. In order to confirm these suspicions, however, more information is needed on the chemical pathways that form CH$_{3}$SH under interstellar conditions. This detection is a first step towards understanding the formation and evolution of sulphur-based COMs, allowing for further comparison into the oxygen and sulphur chemistry for the simplest COMs, and serves as a template for future studies of even more complex sulphur-bearing species. 

\section{Acknowledgments}
H.A.B., S.S., C.P.E., V.L., and P.C. greatly acknowledge the support of the Max Planck Society. J.-C.G. thanks the CNRS national program PCMI (Physics and Chemistry of the Interstellar Medium) and the CNES for a grant (CMISTEP). Part of the research described in this paper was performed at the Canadian Light Source, a national research facility of the University of Saskatchewan, which is supported by the Canada Foundation for Innovation (CFI), the Natural Sciences and Engineering Research Council (NSERC), the Canadian Institutes of Health Research (CIHR), the Government of Saskatchewan, and the University of Saskatchewan. We acknowledge the SOLEIL facility for provision of synchrotron radiation and personnel under the proposal 20230044.

\newpage
\appendix
\restartappendixnumbering

\section{Synthesis of CH$_{2}$DSH} \label{sec:exp}
The synthesis was performed at the Institute of Chemical Sciences in Rennes, France and the final gas production step performed at the Max Planck Institute for Extraterrestrial Physics, Garching, Germany. \\
LiAlD$_{4}$ (1.0 g, 24 mmol) and dry diethyl ether (100 mL) were added to a 250 mL three-necked flask under nitrogen and fitted with a reflux condenser and a dropping funnel. Tributyltin chloride (22.1 g, 68 mmol) was added dropwise and the mixture was heated under reflux for 1 h. After cooling to room temperature, the mixture was carefully added to 30 mL of cold water, the organic phase was separated and the organic compounds were extracted twice with diethyl ether (2 x 50 mL) from the aqueous phase. After drying over MgSO$_{4}$ and evaporation of the solvent, Bu$_{3}$SnD (Tri-n-butyltin deuteride) was distilled in vacuo (bp$_{0.1}$: 80 $^{\circ}$C) and obtained a yield of 90$\%$ (17.7 g). 

Diiodomethane (5.3 g, 20 mmol) and tetraethylene glycol dimethyl ether (15 mL) were added to a 100 mL three-necked flask connected to a vacuum line (0.1 mbar) fitted with two U-traps with stopcocks. The first was immersed in a cold bath at $-$50 $^{\circ}$C and the second in a liquid nitrogen bath. Tributyltin deuteride (6.1 g, 21 mmol) was added to the solution through a septum and the mixture was stirred at room temperature for 4 h. An additional 1.2 g (4.0 mmol) of Tributyltin deuteride was added after 3 h. High-boiling impurities were condensed in the first trap and pure iodomethane-d (CH$_{2}$DI, 2.3 g, 16 mmol) was condensed in the second trap and obtained with a yield of 81$\%$.

NaBH$_{2}$S$_{3}$ (Sulfurated Sodium Borohydride), prepared following the synthesis reported in \cite{synthNa}, was then combined with  CH$_{2}$DI following the synthesis reported by \cite{synth2} to produce a methane-d-thiol salt. The solvent was removed under vacuum for 1 h. Methane-d-thiol salt (3 g) and tetraethylene glycol dimethyl ether (10 mL) were added to a 100 mL double-necked flask connected to a vacuum line fitted with two U-traps with stopcocks, the first cooled to $-$100 $^{\circ}$C and the second immersed in a liquid nitrogen bath. The flask was degassed and octanoic acid (about 200 $\mu$L) was introduced through a septum using a syringe. CH$_{2}$DSH vaporized immediately and was condensed in the second trap, while the high-boiling impurities were trapped in the first trap. The addition of octanoic acid was repeated approximately 6 times every 5-10 minutes. The isotopic purity was higher than 95$\%$, but the sample contained about twice as much hydrogen sulphide.

\newpage
\section{Astronomical Detection}
\begin{figure}[b]
    \centering
	\includegraphics[scale=0.55]{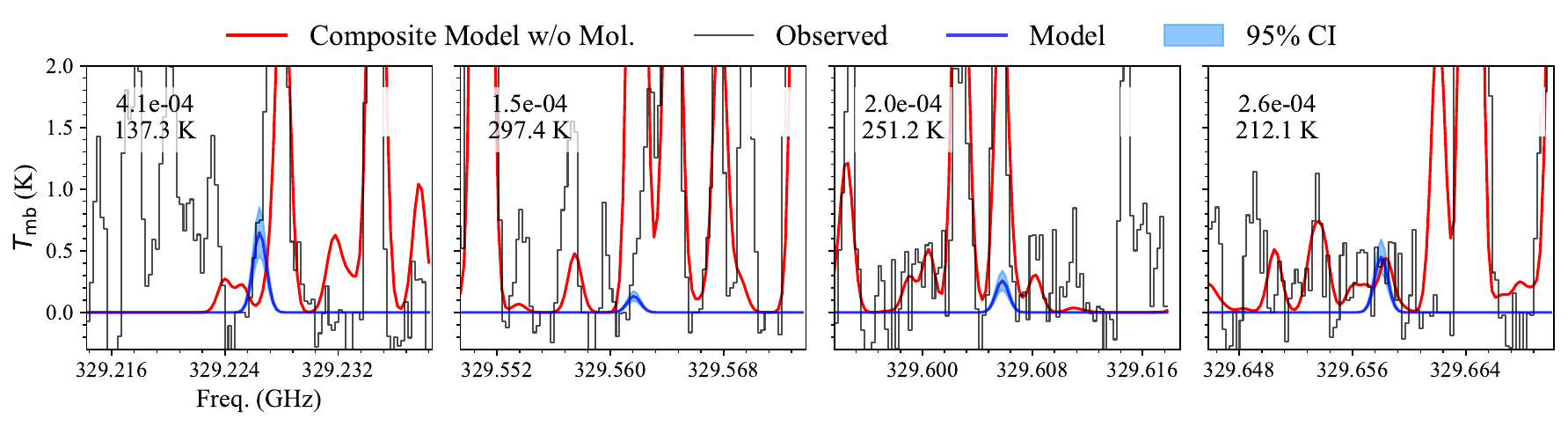}
 \centering 
 \includegraphics[scale=0.55]{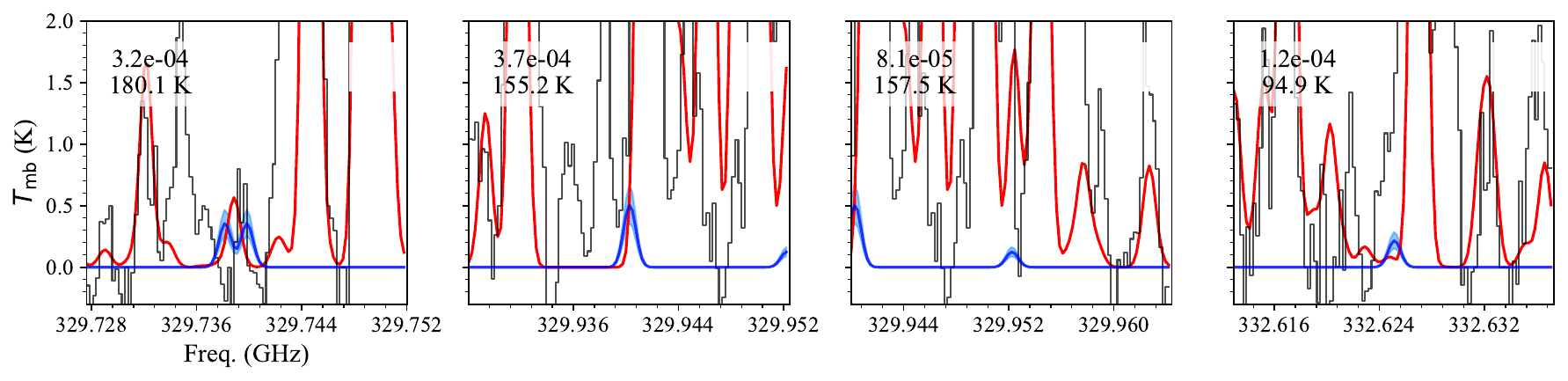}
  \centering 
 \includegraphics[scale=0.55]{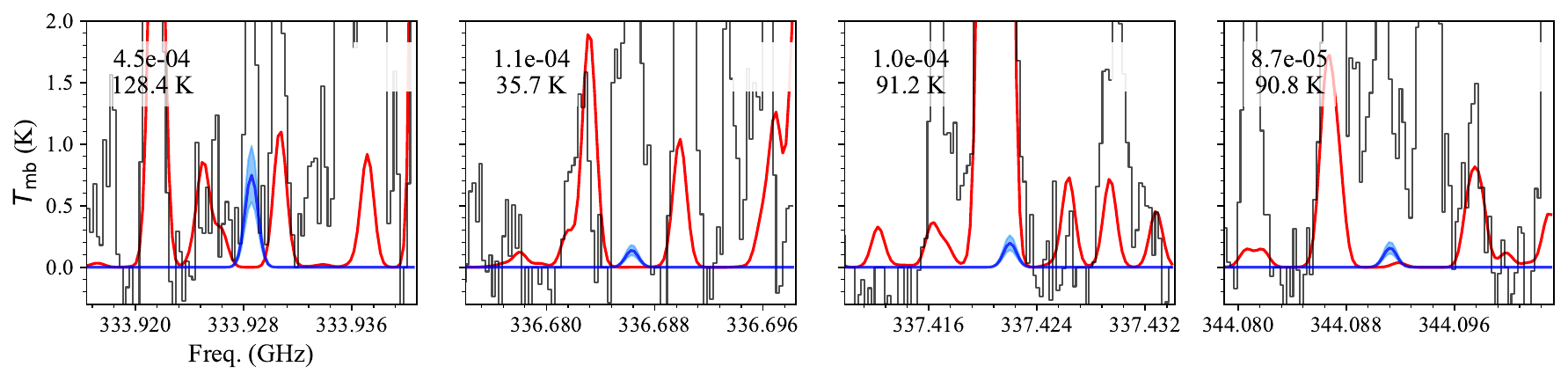}
  \centering 
 \includegraphics[scale=0.55]{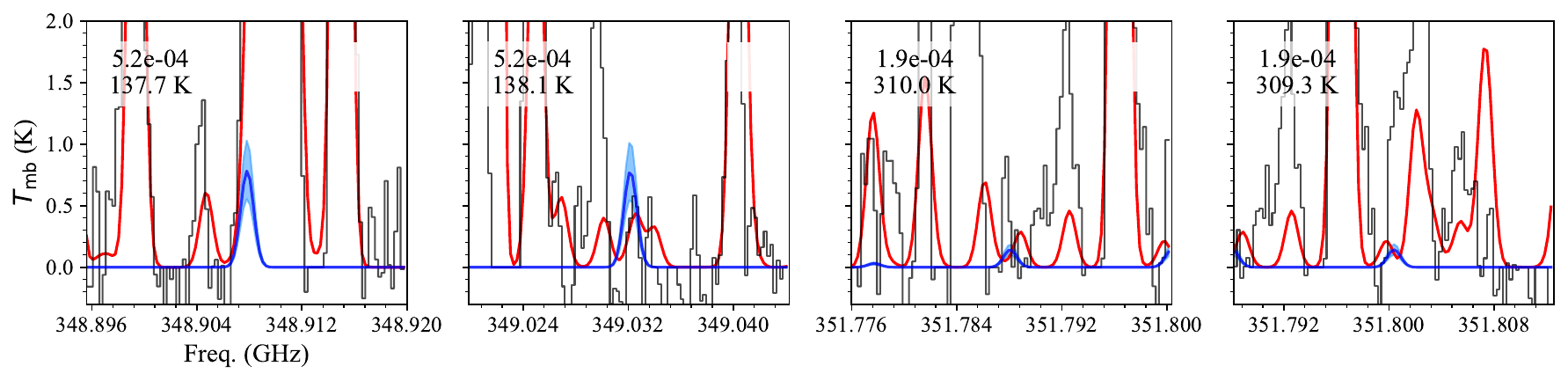}
  \centering 
 \includegraphics[scale=0.55]{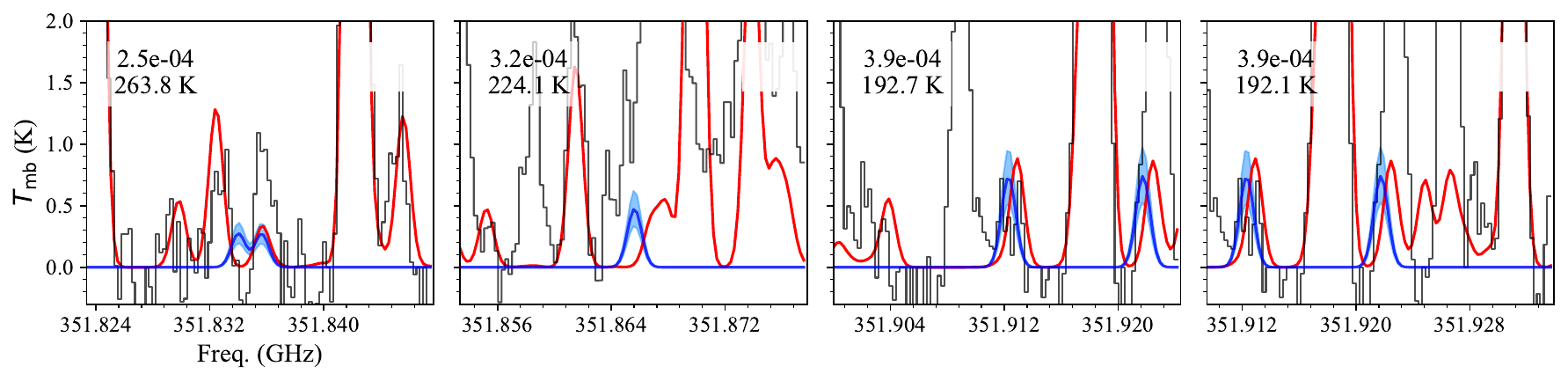}

 \end{figure}
 \begin{figure}
   \centering 
 \includegraphics[scale=0.55]{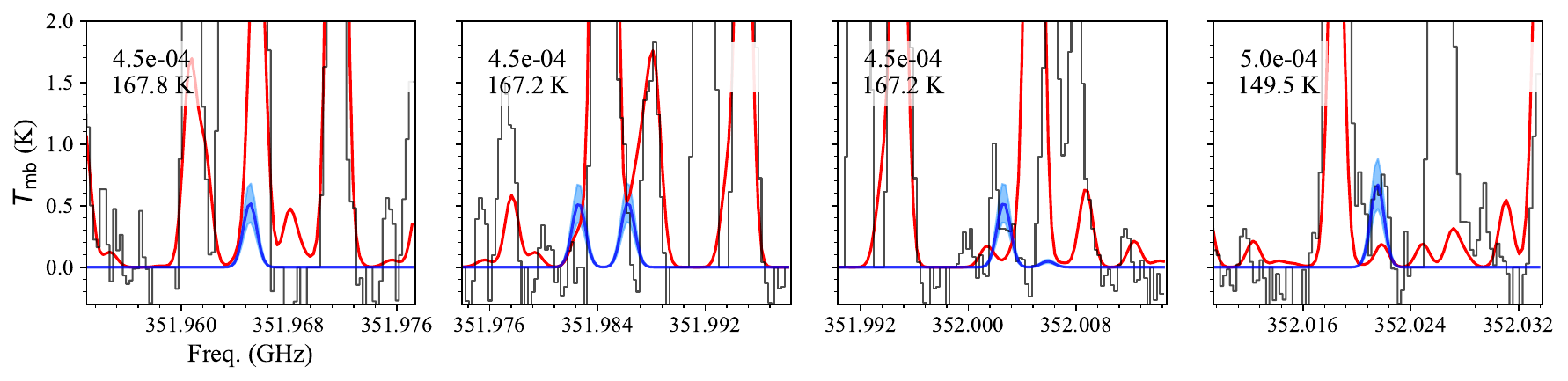}
  \centering 
 \includegraphics[scale=0.55]{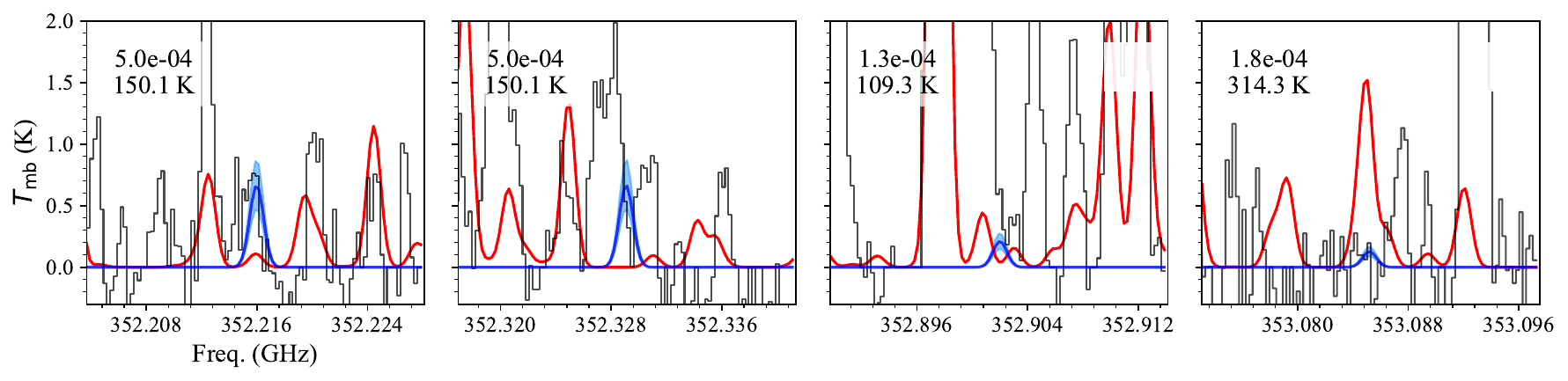}
  \centering 
 \includegraphics[scale=0.55]{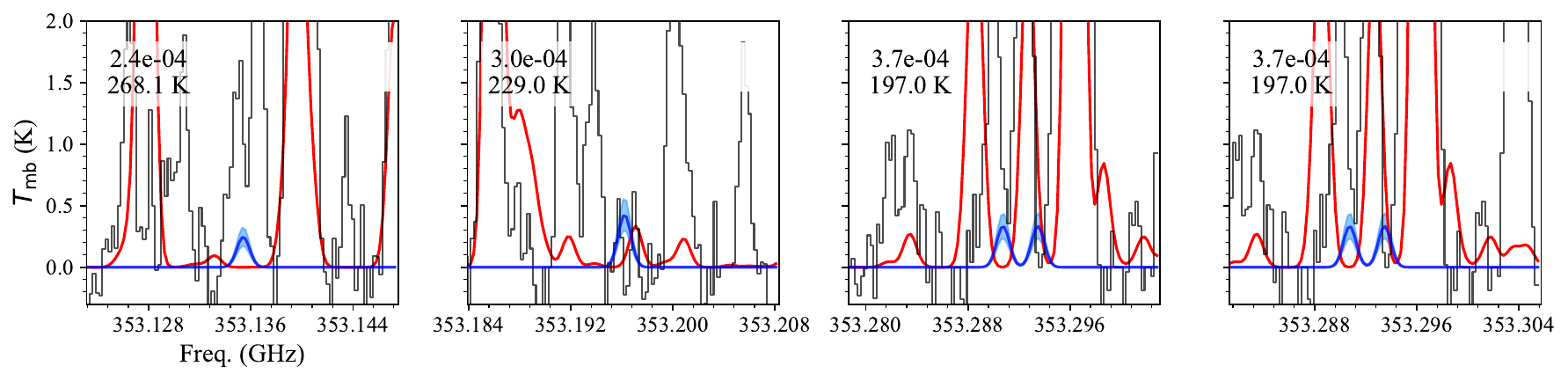}
  \centering 
 \includegraphics[scale=0.55]{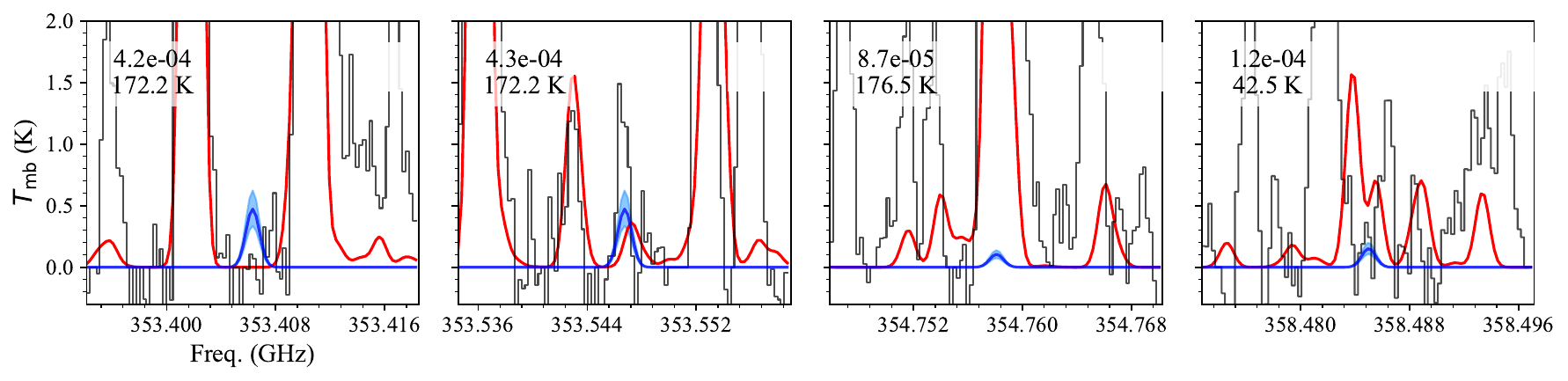}
  \centering 
 \includegraphics[scale=0.55]{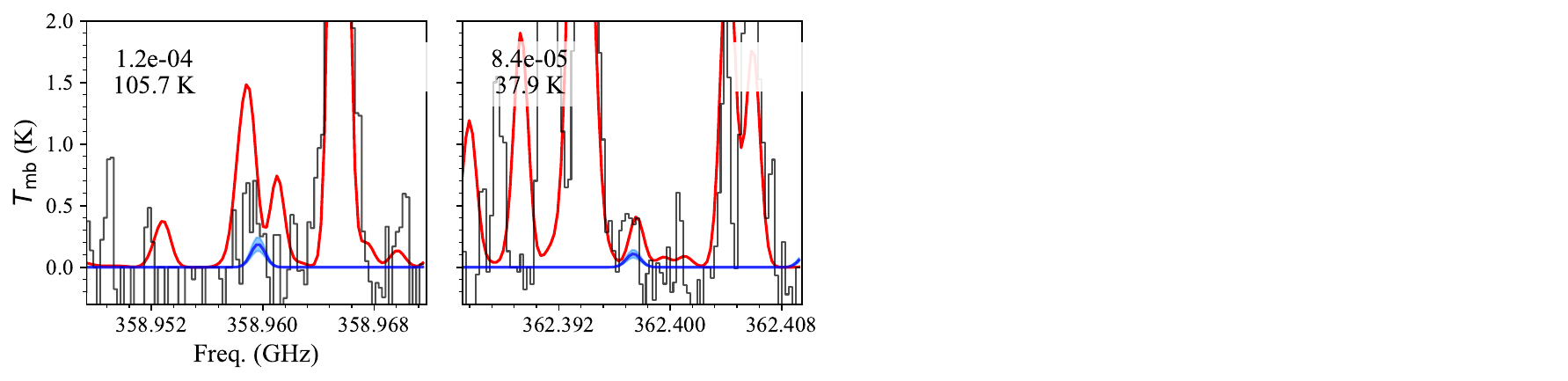}
	\caption{A selection of the detected transitions of CH$_{2}$DSH in IRAS 16293-2422 B, where the black trace is the PILS data, the red is the fit of other known molecules and the blue is the fit to CH$_{2}$DSH. The top left of each panel displays the Einstein A coefficient (in s$^{-1}$) and upper energy level (in K). }
	\label{astrosi}
\end{figure}

\begin{table}
\small
	\caption{Spectroscopic properties of the 46 observed transitions of CH$_{2}$DSH towards IRAS 16293-2422 B}
	\centering
	\begin{tabular}{wl{1.1in}wl{0.9in}wl{0.5in}wl{0.3in}wl{0.9in}}
		 \\
		\hline\hline
        Frequency (MHz) & QN$_{up}$ (\textit{J$_{K,\textit{p},\nu}$}) & E$_{up}$(K) & G$_{up}$ & A ($\times$10$^{-4}$ s$^{-1}$)\\
		\hline
Unblended &  & &  & \\
351113.1 & 15$_{0,1,0}$ & 135.0 & 31 & 5.4\\
351075.6 & 15$_{0,2,1}$ & 135.7 & 31 & 5.4\\
353631.8 & 15$_{1,1,1}$ & 137.7 & 31 & 5.2\\
350411.5 & 15$_{0,1,2}$ & 138.1 & 31 & 5.2\\
357649.5 & 15$_{1,2,2}$ & 192.1 & 31 & 3.9\\
352110.8 & 15$_{2,2,0}$ & 192.7 & 31 & 3.9\\
355325.6 & 15$_{2,1,2}$ & 139.5 & 31 & 5.0\\
351884.1 & 15$_{5,1,0}$/ 15$_{5,2,0}$ & 145.5 & 31 & 5.2\\
Blended &  & &  & \\
 329226.5 & 14$_{2,2,2}$ & 137.3 & 29 & 4.1\\
 329561.7 & 14$_{7,1,2}$/14$_{7,2,2}$ & 297.4 & 29 & 1.5\\
 329605.8 & 14$_{6,1,2}$/14$_{6,2,2}$ & 251.2 & 29 & 2.0\\
 329658.0 & 14$_{5,1,2}$/14$_{5,2,2}$ & 212.1 & 29 & 2.6\\
 329739.8 & 14$_{4,1,2}$ & 180.1 & 29 & 3.2\\
 329940.3 & 14$_{3,2,2}$ & 155.2 & 29 & 3.7\\
 329952.2 & 16$_{0,1,2}$ & 157.5 & 33 & 8.1\\
 332625.1 & 12$_{1,1,2}$ & 94.9 & 25 & 1.2\\ 
 333928.6 & 14$_{1,2,2}$ & 128.4 & 29 & 4.5\\
 336686.3 & 5$_{2,2,2}$ & 35.7 & 11 & 1.1\\
 337422.0 & 12$_{1,2,1}$ & 91.2 & 25 & 1.0\\
 344091.2 & 12$_{1,1,0}$ & 90.8 & 25 & 8.7\\ 
 348907.8 & 15$_{1,1,0}$ & 137.7 & 31 & 5.2\\
 349032.2 & 15$_{1,2,1}$ & 138.1 & 31 & 5.2\\ 
 351788.1 & 15$_{7,1,1}$/15$_{7,2,1}$ & 310.0 & 31 & 1.9\\
 351800.4 & 15$_{7,1,0}$/15$_{7,2,0}$ & 309.3 & 31 & 1.9\\ 
 351835.6 & 15$_{6,1,0}$/15$_{6,2,0}$ & 263.8 & 31 & 2.5\\
 351865.6 & 15$_{5,1,1}$/15$_{5,2,1}$ & 224.1 & 31 & 3.2\\
 351912.3 & 15$_{4,1,0}$/15$_{4,2,0}$ & 192.7  & 31 & 3.9\\
 351921.7 & 15$_{4,1,1}$/15$_{4,2,1}$ & 192.1 & 31 & 3.9\\ 
 351965.1 & 15$_{3,1,0}$ & 167.8 & 31 & 4.5\\
 351986.3 & 15$_{3,2,1}$ & 167.2  & 31 & 4.5\\ 
 352002.6 & 15$_{3,1,1}$ & 167.2 & 31 & 4.5\\
 352021.5 & 15$_{2,1,1}$ & 149.5 & 31 & 5.0\\  
 352216.0 & 15$_{2,2,1}$ & 150.1 & 31 & 5.0\\ 
 352329.1 & 15$_{2,1,0}$ & 150.1 & 31 & 5.0\\
 352902.0 & 13$_{1,1,2}$ & 109.3 & 27 & 1.3\\
 353085.2 & 15$_{7,1,2}$/15$_{7,2,2}$ & 314.3 & 31 & 1.8 \\  
 353135.4 & 15$_{6,1,2}$/15$_{6,2,2}$ & 268.1 & 31 & 2.4\\
 353196.2 & 15$_{5,1,2}$/15$_{5,2,2}$ & 229.0 & 31 & 3.0\\
 353290.7 & 15$_{4,2,2}$ & 197.0 & 31 & 3.7\\
 353293.4 & 15$_{4,1,2}$ & 197.0 & 31 & 3.7\\ 
 353406.3 & 13$_{1,1,2}$ & 172.2 & 31 & 4.2 \\
 353546.8 & 15$_{3,2,2}$ & 172.2 & 31 & 4.3\\
 354758.1 & 17$_{0,1,2}$ & 176.5 & 35 & 8.7 \\
 358485.0 & 6$_{2,2,2}$ & 42.5 & 13 & 1.2\\
 358959.7 & 13$_{1,2,1}$ & 105.7 & 27 & 1.2\\
 362397.4 & 6$_{2,2,1}$ & 37.9 & 13 & 8.4\\
		\hline \hline
		\label{QNs}
	\end{tabular}
\end{table}

\newpage
\bibliography{CH2DSH}{}
\bibliographystyle{aasjournal}

\end{document}